\documentclass{iaus}
\usepackage{graphicx}

\title[LMXBs in GCs and metallicity] 
{Effect of metallicity on low mass X-ray binaries in globular clusters.}

\author[Ivanova]   
{Natalia Ivanova$^1$}

\affiliation{$^1$Physics and Astronomy Department, Northwestern University, 
2145 Sheridan Rd, Evanston IL 60208 \break email: nata@northwestern.edu}

\pubyear{2005}
\volume{230}  
\pagerange{1--2}
\date{19 September, 2005}
\jname{Populations of High Energy Sources in Galaxies}
\editors{Evert J.A. Meurs \& Giuseppina Fabbiano, eds.}
\begin{document}

\maketitle

\begin{abstract}
We propose that the observed difference in the formation rates
of bright low-mass X-ray binaries in metal-rich and 
metal-poor globular clusters can be explained by taking into account 
the difference in the stellar structure of  main sequence  
donors with masses between $\sim 0.85 M_\odot$ and $\sim 1.25 M_\odot$
at different metallicities.
This difference is caused by the absence of 
an outer convective zone in metal-poor main sequence stars in this mass range.
In the result, magnetic braking,  a powerful mechanism of orbital shrinkage, does not operate
and dynamically formed 
main sequence - neutron star binaries fail to
start mass transfer or appear as bright low-mass X-ray binaries.
\end{abstract}
\firstsection 
%
%
\section{Observational motivations}

Bright low-mass X-ray binaries (LMXBs) in globular clusters (GCs) 
of our Galaxy and of M31 preferentially reside in metal-rich GCs 
(Grindlay 1993, Belazzini et al. 1995).
Recent extragalactic observations revealed the same tendency: 
in NGC 4472 (Kundu et al. 2002, Maccarone et al. 2004) 
and in M87 (Jord{\' a}n et al., 2004) 
metal-rich GCs are $3\pm 1$ times as likely to 
contain bright LMXBs as metal-poor GCs.

In detail, from observations of galactic GCs, 
two currently known GC-residing persistent LMXBs  
with a possibly  non-degenerate (but also not a subgiant)  
companion are contained only in metal-rich clusters 
(Verbunt \& Lewin 2005), as well as
two known GC-contained luminous transient LMXBs with 
a possibly  non-degenerate companion.
Five persistent bright LMXBs are observed in 
relatively metal-poor clusters. 
Four of them show indications of being ultracompact 
and one is likely to be a subgiant according 
to its orbital period. 
One luminous transient LMXB in a metal-poor cluster 
is identified as an ultracompact, and
for the bright LMXB in Terzan 1 (metal-poor),  
no information on whether it is ultracompact or not.

This data set hints that possibly 
bright LMXBs consisting of a NS and a 
main sequence (MS) star can preferably be 
formed in metal-rich clusters.
\section{Dynamically formed NS-MS binaries}
NS-MS binaries could form via exchange encounters 
or via tidal captures.
In the first case, the
minimum period of the post-exchange binary 
is limited both by the period at contact for a given eccentricity 
and by the period determined by the binary energy conservation during the encounter. 
Due to the energy conservation, when NS replaces a less massive star, 
the binary separation 
in the formed post-exchange binary is larger than the binary separation
in the pre-exchange binary.

The binary periods of tidally captured binaries are limited at the upper 
limit by the closest approach at which tidal interactions
are still strong enough to make a bound system.
The lower limit is determined by the closest approach 
at which a MS star will pass and still will not be destroyed.
The parameter space for tidally captured binaries where MS star
does not overfill its Roche lobe at the closest approach is very restricted.
\section{Successful LMXBs Candidates}
NS-MS binaries are evolved under the influence 
of gravitational radiation, 
magnetic braking and tides.
In metal-poor clusters, only stars $\le 0.85 M_\odot$ 
have the developed outer convective zone
 -- only there magnetic braking and effective convective tides operate.
This determines the maximum initial binary period, 
for different eccentricities,
such that the binary will start the MT before the MS star leaves the MS 
or another dynamical encounter occurs.

In the result, the parameter-space available for post-exchange 
binaries that can successfully start MT in metal-rich clusters 
is substantially larger than in metal-poor clusters.
Also, in metal-poor clusters,  the efficiency of tidal captures 
is significantly reduced for 
MS stars with radiative envelopes ($\ge 0.85 M_\odot$) and only MS
stars with masses less than $0.55 M_\odot$ can be formed via tidal capture
without overfilling  their Roche lobe during the event.
\section{Persistent or transient?}
In metal-poor clusters, all MT NS-MS binaries with a MS star $\ge 0.85 M_\odot$ will 
evolve on the time-scale predicted by gravitational radiation 
until the donor will decrease its mass enough to develop the
deep outer convective zone. 
Before this moment, MT binary appears as a transient X-ray source.
When the deep outer convective zone is developed, 
MT rates jumps and an LMXB can become persistent
only for a very short period of time.

In metal-rich clusters, an LMXB appears first as a persistent LMXB.
When donor decreases its mass to $\sim 0.7 M_\odot$, 
the MT rate becomes very close to critical and remains 
close for a long time. 
Even slight discrepancy in the value of $\dot M_{\rm crit}$ 
leads to large differences in 
how much time an LMXB spend as persistent $\tau_{\rm pers}$ 
or as a transient system $\tau_{\rm tr}$.
This affects how many qLMXBs and how many bright LMXBs can be present
in metal-rich clusters.

\begin{acknowledgments}
This work is supported by a {\em Chandra}  Theory Award  
to N.\ Ivanova.
\end{acknowledgments}

\end{document}